\documentclass[conference,9pt]{IEEEtran}

\usepackage{waspaa25}

\usepackage{amsmath}
\usepackage{hyperref}
\usepackage{tikz}
\usetikzlibrary{calc,arrows,shapes,positioning,automata,decorations.pathmorphing}
\usepackage{tkz-euclide}
\usepackage[numbib]{tocbibind}
\usepackage{float}
\usepackage{array}
\usepackage{bm}
\usepackage{siunitx}  
\usepackage{balance}

\begin{document}
\renewcommand{\ttdefault}{cmtt}

\tikzset{
block/.style = {draw, fill=white, rectangle, minimum height=3em, minimum width=3em},
tmp/.style  = {coordinate}, 
circ/.style= {draw, fill=white, circle, node distance=1cm, minimum size=0.6cm},
input/.style = {coordinate},
output/.style= {coordinate},
pinstyle/.style = {pin edge={to-,thin,black}.},
snake it/.style={decorate, decoration=snake}
}

\title{RADE: A Neural Codec for Transmitting Speech over HF Radio Channels}

\name{David Rowe\IEEEauthorrefmark{1}, Jean-Marc Valin\IEEEauthorrefmark{2}}
\address{\IEEEauthorrefmark{1}Supported by a grant from Amateur Radio Digital Communications \\
\IEEEauthorrefmark{2}Xiph.Org Foundation }

\maketitle

\begin{abstract}

Speech compression is commonly used to send voice over radio channels in applications such as mobile telephony and two-way push-to-talk (PTT) radio. 
In classical systems, the speech codec is combined with forward error correction, modulation and radio hardware. 
In this paper we describe an autoencoder that replaces many of the traditional signal processing elements with a neural network. 
The encoder takes a vocoder feature set (short term spectrum, pitch, voicing), and produces discrete time, but continuously valued quadrature amplitude modulation (QAM) symbols. 
We use orthogonal frequency domain multiplexing (OFDM) to send and receive these symbols over high frequency (HF) radio channels. 
The decoder converts received QAM symbols to vocoder features suitable for synthesis. 
The autoencoder has been trained to be robust to additive Gaussian noise and multipath channel impairments while simultaneously maintaining a Peak To Average Power Ratio (PAPR) of less than 1~dB. 
Over simulated and real world HF radio channels we have achieved output speech intelligibility that clearly surpasses existing analog and digital radio systems over a range of SNRs.
\end{abstract}

\section{Introduction}

High-frequency (HF) push-to-talk (PTT) radio has the benefit of operating without infrastructure over ranges of several thousands of km.
Applications include humanitarian, remote area, emergency and government communication when access to cellular and satellite systems cannot be guaranteed. 
The HF radio signal propagates from the transmitter to the receiver via reflection from upper layers of the atmosphere.
Typically the signal is reflected multiple times by various sub-layers thus multiple, time shifted versions of the signal arrive and are summed at the receiver (multipath propagation).
Single side band (SSB) -- an analog communications system that was invented in 1915~\cite{4051940} -- entered widespread use in the 1950's~\cite{4051939} and remains the de facto standard on HF due to its power and bandwidth efficiency, and robustness to multipath propagation. 
The voice quality of analog and digital HF speech systems remains low compared to modern cellular and Internet telephony services and has not seen significant improvement in 70 years. 
Key requirements for HF voice services are voice quality, narrow RF bandwidth, low SNR operation, robustness to multipath channels, and one way latency of less than 200~ms \cite{project25} to support PTT speech.

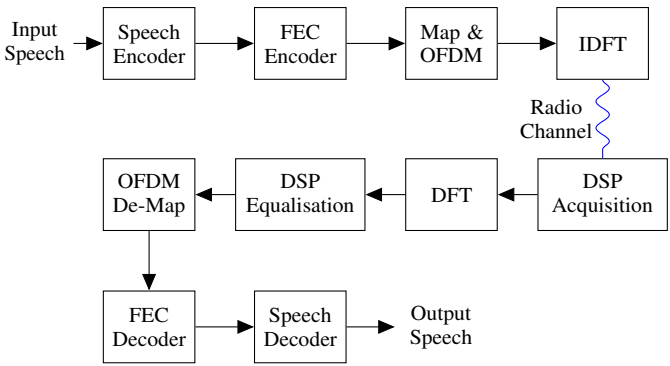
\begin{figure}[h]
\begin{center}
\begin{tikzpicture}[auto, node distance=1.5cm,>=triangle 45,x=1.0cm,y=1.0cm,align=center,text width=1cm,font=\footnotesize]

\node [input] (rinput) {};
\node [block, right of=rinput, node distance=1cm] (speech_enc) {Speech Encoder};
\node [block, right of=speech_enc,node distance=2cm] (fec_enc) {FEC Encoder};
\node [block, right of=fec_enc,node distance=2cm] (map_ofdm) {Map \& OFDM};
\node [block, right of=map_ofdm,node distance=2cm] (idft) {IDFT};
\node [block, below of=idft,node distance=2cm,text width=1.5cm] (dsp_acq) {DSP Acquisition};
\node [block, left of=dsp_acq,node distance=2cm] (dft) {DFT};
\node [block, left of=dft,node distance=2cm,text width=1.5cm] (dsp_eq) {DSP Equalisation};
\node [block, left of=dsp_eq,node distance=2cm] (ofdm_demap) {OFDM De-Map};
\node [block, below of=ofdm_demap,node distance=1.75cm] (fec_dec) {FEC Decoder};
\node [block, right of=fec_dec,node distance=2cm] (speech_dec) {Speech Decoder};
\node [output, right of=speech_dec,node distance=1.25cm] (routput) {};

\draw [->] node[left,text width=0.8cm] {Input Speech} (rinput) -- (speech_enc);
\draw [->] (speech_enc) -- (fec_enc);
\draw [->] (fec_enc) -- (map_ofdm);
\draw [->] (map_ofdm) -- (idft);
\path [draw=blue, snake it] (idft) -- node[left] {Radio Channel} (dsp_acq);
\draw [->] (dsp_acq) -- (dft);
\draw [->] (dft) -- (dsp_eq);
\draw [->] (dsp_eq) -- (ofdm_demap);
\draw [->] (ofdm_demap) -- (fec_dec);
\draw [->] (fec_dec) -- (speech_dec);
\draw [->] (speech_dec) -- (routput) node[right, text width=1cm] {Output Speech};

\end{tikzpicture}
\end{center}
\caption{Classical DSP speech over radio system employing OFDM modulation.}
\label{fig:classical_block}
\end{figure}

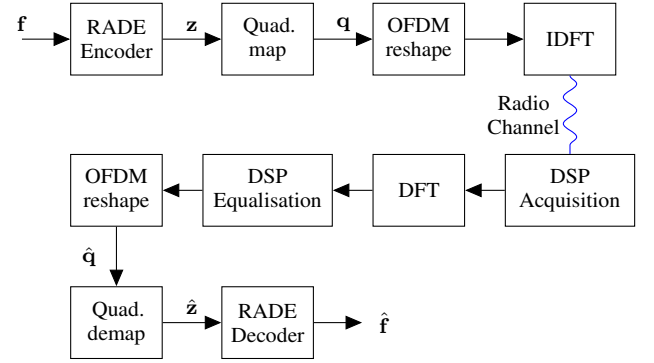
\begin{figure}[h]
\begin{center}
\begin{tikzpicture}[auto, node distance=2cm,>=triangle 45,x=1.0cm,y=1.0cm,align=center,text width=1cm, font=\footnotesize]

\node [input] (rinput) {};
\node [block, right of=rinput,node distance=1.25cm] (rade_enc) {RADE Encoder};
\node [block, right of=rade_enc,node distance=2cm] (quad_map) {Quad. map};
\node [block, right of=quad_map,node distance=2cm] (ofdm_frame) {OFDM reshape};
\node [block, right of=ofdm_frame,node distance=2cm] (idft) {IDFT};
\node [block, below of=idft,node distance=2cm,text width=1.5cm] (dsp_acq) {DSP Acquisition};
\node [block, left of=dsp_acq,node distance=2cm] (dft) {DFT};
\node [block, left of=dft,node distance=2cm,text width=1.5cm] (dsp_eq) {DSP Equalisation};
\node [block, left of=dsp_eq,node distance=2cm] (ofdm_deframe) {OFDM reshape};
\node [block, below of=ofdm_deframe,node distance=1.75cm] (quad_demap) {Quad. demap};
\node [block, right of=quad_demap,node distance=2cm] (rade_dec) {RADE Decoder};
\node [output, right of=rade_dec,node distance=1.25cm] (routput) {};

\draw [->] node[above,text width=1cm] {$\bf{f}$} (rinput) -- (rade_enc);
\draw [->] (rade_enc) -- node[above, text width=0.5cm] {$\bf{z}$} (quad_map);
\draw [->] (quad_map) -- node[above, text width=0.5cm] {$\bf{q}$} (ofdm_frame); 
\draw [->] (ofdm_frame)-- (idft); 
\path [draw=blue, snake it] (idft) -- node[left] {Radio Channel} (dsp_acq);
\draw [->] (dsp_acq)-- (dft); 
\draw [->] (dft)-- (dsp_eq);
\draw [->] (dsp_eq)-- (ofdm_deframe); 
\draw [->] (ofdm_deframe)--  node[left, text width=0.5cm] {$\hat{\bf{q}}$} (quad_demap); 
\draw [->] (quad_demap) -- node[above, text width=0.5cm] {$\hat{\bf{z}}$} (rade_dec); 
\draw [->] (rade_dec) -- node[right, text width=1cm] {$\hat{\bf{f}}$} (routput) ;

\end{tikzpicture}
\end{center}
\caption{RADE system, which employs ML combined with OFDM and classical DSP synchronisation.
The features $\mathbf{f}$ are obtained from input speech using a feature extractor, and output speech is synthesised from $\hat{\bf{f}}$ using the FARGAN vocoder.
The features $\mathbf{f}$ are updated at 100~Hz, the latent vector $\mathbf{z}$ at 25~Hz, and the sample rate over the channel is $F_s=8000$~Hz.}
\label{fig:ml_dsp_framework}
\end{figure}

Consider the classical DSP speech over radio system in Fig.~\ref{fig:classical_block}. 
Speech samples from a microphone are compressed to a low bit rate using a speech encoder. 
Forward error correction (FEC) adds redundant bits to protect the sensitive payload speech bits from channel errors. 
The bitstream is then converted to a signal suitable for transmission over the radio channel using a modulator (e.g. a sequence of QAM symbols mapped to OFDM sub-carriers). 
The received radio signal is demodulated to a bit stream; the FEC decoder attempts to correct any bit errors, and the speech decoder converts the signal back into a sampled speech signal for replay over a loudspeaker.
Classical DSP digital voice systems have the following drawbacks: separating the speech coding and channel protection (FEC) leads to inefficiencies; the use of largely linear DSP means the inability to exploit non-linear dependencies when compressing voice; difficulty in handling multipath channels; long latencies due to use of interleavers/FEC to overcome multipath fading; and low quality speech from low bit rate classical DSP vocoders.
They exhibit a threshold SNR where the system ceases to work, and speech quality does not gracefully scale with available channel SNR (although step changes are possible with mode switching).

This paper proposes RADE (Fig.~\ref{fig:ml_dsp_framework}), a RADio autoEncoder~\cite{oshea2017introductiondeeplearningphysical} designed to efficiently transmit speech over HF radio channels inspired by the RDO-VAE structure from DRED~\cite{valin2024dred}.

\begin{figure}[h]
\begin{center}
\scalebox{.9}{\input {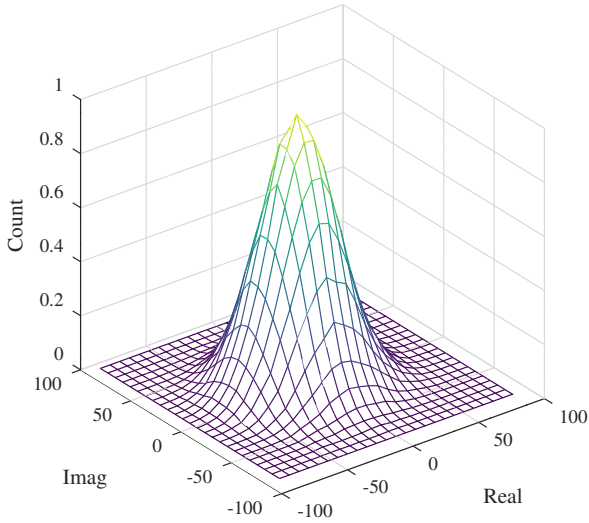}}
\end{center}
\caption{2D histogram of complex RADE encoder QAM symbols with all elements of $\mathbf{z}$ superimposed for a 60 second sample containing multiple speakers.
The bin count has been scaled to a maximum of 1.
Compared to digital QAM constellations (which would appear as Dirac functions on this plot), the RADE constellation looks like noise.
Plotting individual elements of $\mathbf{z}$ produces a similar histogram (no obvious structure).}
\label{fig:histogram}
\end{figure}

The use of ML combines the operations of quantisation, channel encoding and modulation, and allows joint training of the entire end-to-end system to minimise distortion when subjected to the HF channel impairments~\cite{chen2021hybrid,bokaei2025low,yang2022ofdm}.
The powerful non-linear transforms and prediction available in ML allow us to more efficiently model the speech signal and time based evolution of the channel to improve robustness. 
We employ the FARGAN vocoder~\cite{valin2024low} for high quality neural speech synthesis, however our work is applicable to any neural and even classical vocoders with a similar feature set.
\clearpage
Our contributions in this paper are:
\begin{enumerate}
\item An autoencoder that combines the classical DSP functions of quantisation, channel coding, and modulation to generate discrete time but continuously valued (analog) QAM symbols directly from vocoder features. 
Unlike classical approaches there is no intermediate bit stream, and the QAM symbols (Fig.~\ref{fig:histogram}) emerge from the training process rather than being members of a well defined, discrete constellation.
\item A training procedure that minimises the end-to-end distortion of vocoder features in the presence of additive Gaussian noise and frequency-selective fading, while simultaneously generating an OFDM waveform with low Peak To Average Power Ratio (PAPR).
\end{enumerate}

In Section~\ref{sec:rade_design} we describe how we have combined a neural encoder and decoder with OFDM to develop the RADE system. 
Training over HF channels at low PAPR is discussed in Section~\ref{sec:training}. 
For testing we have adopted an automatic speech recognition (ASR) based approach over simulated HF channels which we describe in Section~\ref{sec:rade_simulation}, along with an over the air demonstration. 

\section{RADE Design}
\label{sec:rade_design}

Rather than operating directly on the time-domain signal like recent end-to-end neural codecs~\cite{zeghidour2021soundstream}, RADE uses classical acoustic features similar to those used in LPCNet~\cite{valin2019lpcnet}. We use 20-dimensional feature vectors $\mathbf{f}$ that consist of 18~Bark-scale cepstral coefficients, the pitch period, and a voicing parameter. Using classical features not only reduces complexity, but makes it easy to replace the vocoder without breaking compatibility. Moreover, we show in Section~\ref{sec:rade_simulation} that our choice of features is not a limiting factor for our application.

The RADE encoder and decoder form an autoencoder that is trained to minimise the reconstruction loss $\mathcal{L}(\mathbf{f}, \hat{\mathbf{f}})$ between the input features~$\mathbf{f}$ and the decoded features~$\hat{\mathbf{f}}$, as defined in Eq.~(12) of~\cite{valin2024dred}. The feature vectors $\mathbf{f}_n$ generated every 10~ms are concatenated and passed to the RADE encoder every 40~ms (frame rate of 25~Hz). They are transformed to a $d=80$ dimensional vector $\bf{z}$ by a stack of 1D convolutional (conv) and gated recurrent units (GRU), arranged in a DenseNet-like~\cite{huang2017densely} topology.
This encoder was derived from RDO-VAE~\cite{valin2024dred}, with the quantisation steps deleted. 
For 40~ms time-step $n=0,4,8,...$, each stage can be represented by:
\begin{equation}
\mathbf{z}_{i+1} = [\mathbf{z}_i,\mathrm{conv_i}([\mathbf{z}_i, \; \mathrm{GRU_i}(\mathbf{z}_i)])], \quad i=1..6
\end{equation}
where $\mathbf{z}_1=\mathrm{dense}([\mathbf{f}_n,...,\mathbf{f}_{n+3}])$, and the encoder output $\mathbf{z}=\mathrm{dense}(\mathbf{z}_6)$.
The decoder has a similar, but symmetrical design, but includes an additional gated linear unit (GLU) in each stage.
\begin{equation}
\mathbf{y}_{i+1} = [\mathbf{y}_i,\mathrm{conv_i}([\mathbf{y}_i, \; \mathrm{GLU_i}(\mathrm{GRU_i}(\mathbf{y}_i))])], \quad i=1..6
\end{equation}
where $\mathbf{y}_1=\mathrm{dense}(\mathbf{\hat{z}})$ and the output feature vectors $[\mathbf{\hat{f}}_n,...,\mathbf{\hat{f}}_{n+3}]=\mathrm{dense}(\bf{y}_6)$.

For bandwidth efficient transmission over the channel the elements of $\mathbf{z}$ are mapped to $d/2=40$ complex QAM symbols $\mathbf{q}$. 
Compared to classical digital modulation, the elements of $\mathbf{q}$ can be viewed as continuously valued (analog) QAM symbols.

For transmission over the HF multipath channel we employ OFDM with pilots symbols. 
We reshape the serial stream of QAM symbols at rate $R_q$ as $N_c$ parallel sub-carriers, each running at a symbol rate of $R_s=R_q/N_c$ symbols/s, where $R_s$ is chosen based on delay spread considerations. 
We have chosen $N_c=30$ and $R_s=50$~Hz. 

This hybrid ML-DSP design has several benefits:
\begin{enumerate}
\item OFDM performs equalisation using a single complex multiply of each symbol which allows us to efficiently represent the multipath channel fading in the ML frame-rate processing. 
 
\item The $F_s=8000$~Hz sample rate processing is performed efficiently in classical DSP, with the ML processing at a much slower frame rate $R_f=25$~Hz.
\item We perform acquisition, synchronisation, and sample rate conversion using well-known DSP techniques. Alternatively, these tasks can also be accomplished using ML \cite{dorner2017deep}\cite{felix2018ofdm}.
\end{enumerate}
A disadvantage of OFDM is high peak to average power ratio (PAPR), which for a given transmitter peak power reduces the available SNR at the receiver. 
We have largely overcome this issue via training, as explained in Section~\ref{sec:training}.
 
\begin{figure}
\begin{center}
\begin{tikzpicture}[>=triangle 45,x=1.0cm,y=1.0cm,font=\footnotesize]
\draw[->] (-0.5,-0.5) -- (7,-0.5) node [below, align=left, text width=2cm]{Time};
\draw[->] (-0.5,-0.5) -- (-0.5,4) node [above]{Freq};

\def\step{0.7}
\def\xlines{8}

\def\xmin{0.5}

\def\xminA{1.5}
\def\xmaxA{4.5}
\def\ymin{2}
\def\ymax{4}
\def\ymaxA{3}

\foreach \i in {0,...,\xlines} {
	\draw [very thin,gray] (\xmin+\i*\step,2*\step) -- (\xmin+\i*\step,4*\step);
}

\foreach \i in {2,...,4} {
	\draw [very thin,gray] (\xmin,\i*\step) -- (\xmin+\xlines*\step,\i*\step);
}

\foreach \i in {2,...,3} {
	\foreach \j in {1,6} {
		\draw (\xmin+\j*\step+0.5*\step,\i*\step+0.5*\step) node [red] {P};
	}
}
\foreach \i in {2,...,3} {
	\foreach \j in {0,2,3,4,5,7} {
		\draw (\xmin+\j*\step+0.5*\step,\i*\step+0.5*\step) node [blue] {D};
	}
}

\foreach \i in {0,...,\xlines} {
	\draw [very thin,gray] (\xmin+\i*\step,0) -- (\xmin+\i*\step,\step);
}
\foreach \i in {0,...,1} {
	\draw [very thin,gray] (\xmin,\i*\step) -- (\xmin+\xlines*\step,\i*\step);
}

\foreach \j in {1,6} {
	\draw (\xmin+\j*\step+0.5*\step,0.5*\step) node [red] {P};
}
\foreach \j in {0,2,3,4,5,7} {
	\draw (\xmin+\j*\step+0.5*\step,0.5*\step) node [blue] {D};
}

\draw [dashed] (\xmin+3.5*\step,1*\step) -- (\xmin+3.5*\step,2*\step);
\foreach \i in {0,2,3} {
    \draw [dashed] (\xmin-\step,\i*\step+0.5*\step) -- (\xmin,\i*\step+0.5*\step);
    \draw [dashed] (\xmin+\xlines*\step,\i*\step+0.5*\step) -- (\xmin+\xlines*\step+\step,\i*\step+0.5*\step);
}

\def\ncx{9.5}
\draw [very thin,gray] (\xmin+\ncx*\step,4*\step) -- (\xmin+\ncx*\step+\step,4*\step);
\draw [very thin,gray] (\xmin+\ncx*\step,0) -- (\xmin+\ncx*\step+\step,0);
\draw [<->] (\xmin+\ncx*\step+0.5*\step,0) --  node[left] {$N_c$} (\xmin+\ncx*\step+0.5*\step,4*\step);


\def\xa{1}
\def\ya{3.5}
\def\xb{5}

\draw (\xa,\ya) -- (\xb,\ya) -- (\xb,\ya+\step) -- (\xa,\ya+\step) -- (\xa,\ya);
\draw (2,\ya) -- (2,\ya+\step);
\draw [very thin,gray,->] (\xmin+\step,4*\step) -- (\xa,\ya);
\draw [very thin,gray,->] (\xmin+2*\step,4*\step) -- (\xb,\ya);
\node[align=center] at (1.5,\ya+0.5*\step) {$CP$};
\node[align=center] at (3.5,\ya+0.5*\step) {$D^\prime$};
\draw [<->] (\xa,\ya+1.5*\step) -- node[above] {$T_{cp}$} (\xa+1,\ya+1.5*\step);
\draw [<->] (\xa+1,\ya+1.5*\step) -- node[above] {$T^\prime_s$} (\xb,\ya+1.5*\step);
\draw [<->] (\xa,\ya+2.5*\step) -- node[above] {$T_s$} (\xb,\ya+2.5*\step); 
\draw [very thin,gray] (\xa,\ya+\step) -- (\xa,\ya+3*\step);
\draw [very thin,gray] (\xb,\ya+\step) -- (\xb,\ya+3*\step);
\end{tikzpicture}
\end{center}
\caption{OFDM modem frame, $P$ denotes pilot symbols, $D$ payload symbols. 
In each frame we have $N_s=4$ payload symbols, and $N_c=30$ carriers. 
Each symbol is comprised of a $T_{cp}=0.004$ second Cyclic Prefix and $T_s^\prime=0.020$ second symbol $D^\prime$.}
\label{fig:ofdm_frame}
\end{figure}
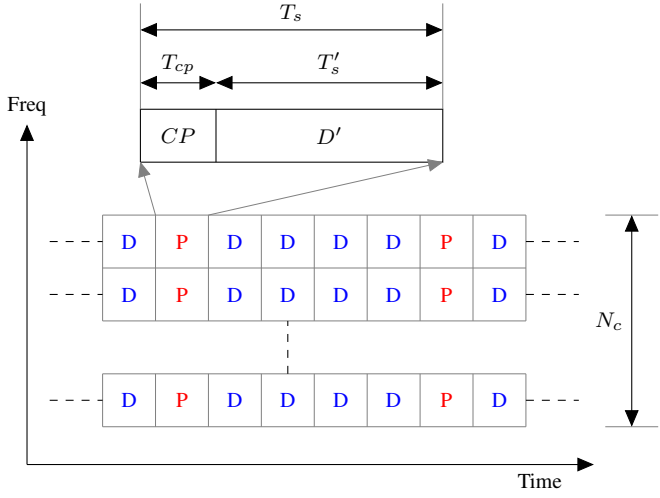

The OFDM frame is shown in Fig.~\ref{fig:ofdm_frame}. 
Pilot symbols are periodically inserted into each OFDM carrier. 
After the IDFT stage, we insert a cyclic prefix to guard against inter-symbol interference.
To achieve an efficient ratio of pilots to data symbols, we place the QAM symbols from three consecutive $\mathbf{z}$ vectors (120 complex QAM symbols) in each OFDM frame, leading to an OFDM frame duration (and algorithmic delay) of 120~ms.

At the receiver the  pilots are used to estimate the time-varying phase of the channel (equalisation), and for initial acquisition (coarse frequency, frame sync) of the received signal.
Phase equalisation also allows small frequency offsets ($\pm2$~Hz) to be handled and frequency drift tracked. 
Neural networks are sensitive to magnitude scaling, so we also use the pilot symbols for coarse magnitude equalisation (gain control) of the received $\mathbf{\hat{z}}$ signal.

The insertion of pilot symbols and the cyclic prefix consume 2~dB of carrier power that would otherwise be available for payload symbols, and require the symbol rate and hence overall RF bandwidth to be increased by 500~Hz to maintain the payload symbol rate. 
An additional 2-dB penalty is incurred due to non-ideal performance of the least squares estimation algorithm used for phase equalisation at our low SNR operating point.
The total cost of the equalization and cyclic prefix insertion is therefore 4~dB. The resulting waveform has an RF bandwidth of approximately 1500~Hz.

\section{Training}
\label{sec:training}

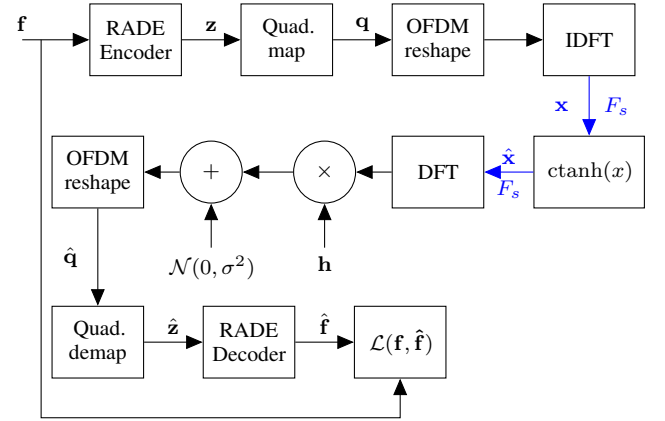
\begin{figure}[h]
\begin{center}
\begin{tikzpicture}[auto, node distance=2cm,>=triangle 45,x=1.0cm,y=1.0cm,align=center,text width=1cm, font=\footnotesize]

\node [input] (rinput) {};
\node [tmp, right of=rinput,node distance=0.25cm] (z) {};
\node [input, right of=rinput,node distance=0.25cm] (junc) {};
\node [block, right of=rinput,node distance=1.5cm] (rade_enc) {RADE Encoder};
\node [block, right of=rade_enc,node distance=2cm] (quad_map) {Quad. map};
\node [block, right of=quad_map,node distance=2cm] (ofdm_frame) {OFDM reshape};
\node [block, right of=ofdm_frame,node distance=2cm] (idft) {IDFT};
\node [block, below of=idft,node distance=1.75cm, text width=1.25cm] (bottleneck) {$\mathrm{ctanh}(x)$};
\node [block, left of=bottleneck,node distance=2cm] (dft) {DFT};

\node [circ, left of=dft,node distance=1.5cm, text width=0.5cm] (multipath) {$\times$};
\node [input, below of=multipath,node distance=1cm] (H) {};
\node [circ, left of=multipath,node distance=1.5cm, text width=0.5cm] (awgn) {$+$};
\node [input, below of=awgn,node distance=1cm] (noise) {};

\node [block, left of=awgn,node distance=1.5cm] (ofdm_deframe) {OFDM reshape};
\node [block, below of=ofdm_deframe,node distance=2.25cm] (quad_demap) {Quad. demap};
\node [block, right of=quad_demap,node distance=2cm] (rade_dec) {RADE Decoder};
\node [block, right of=rade_dec,node distance=2cm] (loss) {$\mathcal{L}(\bf{f},\hat{\bf{f}})$};
\node [tmp, below of=loss,node distance=1cm] (z1) {};

\draw [->] node[above,text width=1cm] {$\bf{f}$} (rinput) -- (rade_enc);
\draw [->] (rade_enc) -- node[above, text width=0.5cm] {$\bf{z}$} (quad_map);
\draw [->] (quad_map) -- node[above, text width=0.5cm] {$\bf{q}$} (ofdm_frame); 
\draw [->] (ofdm_frame)-- (idft); 
\draw [blue,->] (idft) -- node[right, text width=0.5cm] {$F_s$} node[left, text width=0.5cm] {$\bf{x}$} (bottleneck); 
\draw [blue,->] (bottleneck) --  node[below] {$F_s$} node[above] {$\hat{\bf{x}}$} (dft);
\draw [->] (H) node[below,text width=0.5cm] {$\bf{h}$} -- (multipath); 
\draw [->] (dft)-- (multipath); 
\draw [->] (noise) node[below,text width=1.2cm] {$\mathcal{N}(0,\sigma^2)$} -- (awgn); 
\draw [->] (multipath)-- (awgn); 
\draw [->] (awgn)-- (ofdm_deframe); 
\draw [->] (ofdm_deframe)--  node[left, text width=0.5cm] {$\hat{\bf{q}}$} (quad_demap); 
\draw [->] (quad_demap) -- node[above, text width=0.5cm] {$\hat{\bf{z}}$} (rade_dec); 
\draw [->] (rade_dec) -- node[above, text width=1cm] {$\hat{\bf{f}}$} (loss) ;
\draw [->] (z) |- (z1) -- (loss);

\end{tikzpicture}
\end{center}
\caption{Configuration used for training.
A mixed sample rate model is used for joint PAPR minimisation and optimisation for multipath channels.
The rate $F_s$ signals are blue, all other signals are rate $R_s$.
The channel model comprised of $\bf{h}$ and $\mathcal{N}(0,\sigma^2)$ is applied only during training.
}
\label{fig:rade_training}
\end{figure}

Fig.~\ref{fig:rade_training} illustrates the configuration used for training. 
We use a mixed sample rate design, with most of the signal processing occurring at the subcarrier rate $R_s$, and selected portions at the sample rate $F_s$. 
Only the RADE encoder and decoder have trainable parameters. 
The bottleneck is defined as:
\begin{equation}
\mathrm{ctanh(x)} = \tanh(|x|)e^{j\arg[x]} 
\end{equation} 
which simulates the saturation of a transmitter power amplifier by compressing the magnitude but retaining the phase.
We reasoned that a bottleneck applied to the magnitude of the complex time domain signal would encourage the network to maximise the RMS power (and hence minimse the PAPR) given the channel noise and peak power constraint of the bottleneck. Other approaches that include PAPR in the loss function \cite{huleihel2020low} are also possible.

To train we need to apply the $\mathrm{ctanh}(x)$ bottleneck in the time domain, and simultaneously apply a multipath channel model. 
Applying the multipath model in the time domain introduces phase rotations and inter-symbol interference (ISI), which would then require equalisation and removal inside the training loop. 
While possible this would significantly slow down the training process, and require training at the higher rate $F_s$ sample rate ($F_s/R_s=160$ in our design).

We use the simple equalisation properties of OFDM and perform multipath and AWGN channel simulation in the frequency domain.
We assume phase equalisation and ISI removal is performed by the classical DSP stages of the receiver and ignore these these steps during training. 
This reduces multipath channel simulation to magnitude-only fading applied to each frequency domain QAM symbol via a single real-complex multiplication. 

The mixed-rate training system works as follows. 
The transmit QAM symbols are transformed to the time domain with an inverse DFT, the bottleneck applied, then immediately transformed back to the frequency domain. 
The real valued multipath model magnitude samples $\bf{h}$ are applied to the rate $R_s$ (frequency domain) QAM symbols via a simple multiplication.
They are derived from a two path Watterson model~\cite{itu1487}:
\begin{equation}
\label{eq:watterson}
y(t) = x(t)G_1(t) + x(t-d)G_2(t)
\end{equation}
where $x(t)$ is the time domain signal from the transmitter, and $y(t)$ is the output of the multipath fading model.
$G_1$ and $G_2$ are two band-limited complex Gaussian signals with \emph{Doppler Spread} bandwidth $B_{ds}$, and $d$ is the delay spread (path delay) in seconds.
Typically, $B_{ds} \approx 1$~Hz, therefore $G_1$ and $G_2$ slowly vary in amplitude and phase, modelling reflection of the transmitted signal from separate layers of the ionosphere.
The sum of the two terms of (\ref{eq:watterson}) causes notches separated by $1/d$ to appear in the simulated channel, with the position and depth of the notches varying as $G_1$ and $G_2$ evolve.
As $B_{ds}<<R_s$ the channel can be considered stationary over the period of one symbol.
By taking the $z$-transform and evaluating at the centre of each carrier frequency $\omega_c$, the elements $h_c$ of $\bf{h}$ can be computed as:
\begin{equation}
\label{eq:watterson_z}
h_c = |H(e^{j \omega_c})| = |G_1+e^{-j \omega_c d F_s}G_2|
\end{equation}
We train with a delay spread $t=2$~ms and Doppler spreading bandwidth of $B_{ds}=1$~Hz which we denote a multipath poor (MPP) channel.
$G_1$ and $G_2$ are sampled and $\bf{h}$ updated at rate $R_s$.

We use the same 205\nobreakdash-hour training set as~\cite{valin2024dred}, which includes more than 900~speakers in 34~languages and dialects.
It is reshaped into 4\nobreakdash-second sequences, with the AWGN noise for each sequence chosen at random over a 20~dB range $-3<E_q/N_0<17$~dB to encourage operation at a range of SNRs, where $E_q$ is the energy of each QAM symbol, and $N_0$~is the noise power per unit bandwidth.
Given a symbol magnitude $A_q$, the total RMS noise summed across the real an imaginary components can be computed as $\sigma = A_q/\sqrt{E_q/N_0}$.
Training using this model resulted in signals with a PAPR of less that 1~dB and intelligible speech down to $E_q/N_0=-3$~dB on AWGN channels.

The system is trained without pilot symbols or cyclic prefix insertion. 
After training, classical DSP techniques for equalisation and acquisition illustrated in Fig.~\ref{fig:ml_dsp_framework} are wrapped around the core ML to develop the practical, rate $F_s$ speech over HF system described in Section~\ref{sec:rade_design}.

\section{Evaluation and Results}
\label{sec:rade_simulation}

\begin{figure}[h]
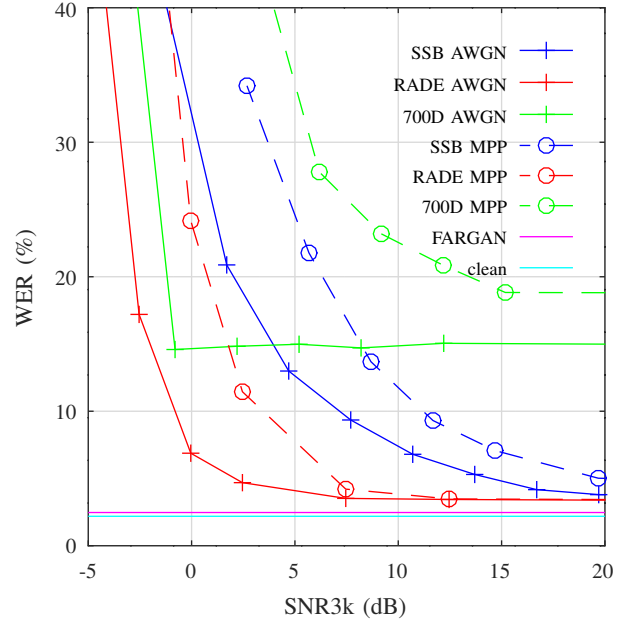

\begin{center}
\input wer_snr.tex
\end{center}
\caption{Word error rate \% versus SNR for simulated AWGN and MPP channels, with clean and FARGAN with clean features as controls. The label SNR3k denotes all signals normalised to a 3000~Hz noise bandwidth.}
\label{fig:wer_snr}
\end{figure}

Informal listening tests on simulated and over the air samples suggest RADE significantly outperforms SSB. 
Since intelligibility -- more than quality -- is the primary goal for HF radio, we use Automatic Speech Recognition (ASR) to evaluate the performance of the proposed system.
Five hundred samples from the Librispeech dataset were passed through RADE, SSB, and FreeDV 700D simulations at a range of SNRs, then post processed by the Whisper ASR system~\cite{radford2022robustspeechrecognitionlargescale}, and the Word Error Rate (WER) measured (Fig.~\ref{fig:wer_snr}).
SSB was simulated by band limiting the input speech to 300-2700~Hz, and applying Hilbert compression such that the mean PAPR was around 8~dB.
FreeDV 700D is an open-source HF digital voice protocol using an OFDM modem, rate $1/2$ FEC, and the Codec 2 classical vocoder~\cite{codec2alg}.
The MPP channel was simulated using the time domain Watterson model Eq. (\ref{eq:watterson}).
The Librispeech speech and Watterson model $G_1$ and $G_2$ datasets used for the evaluation were not part of the training dataset.

We consider two thresholds (a) Link closure - the point where barely intelligible speech can be sent over the system and (b) Good intelligibility, effortless communication.
At the 30\% WER level (link closure), the ASR results indicate a 4~dB improvement for RADE over SSB for both AWGN and MPP channels.
At the 5\% WER level (good quality), the improvement is 13~dB for both channels.
FreeDV 700D exhibits low speech quality with the Librispeech dataset, although we note it is in common use on HF by trained operators.
Similarly, skilled operators can use SSB down to 0~dB SNR.
Note the sharp knee in the FreeDV 700D AWGN curve at $-2$~dB SNR, common in digital speech system due to the abrupt breakdown of the FEC.
RADE and SSB have a more desirable gradual trade off between SNR and speech quality.

These results are based on the SNR at the receiver, and exclude the potential PAPR improvement of 7~dB.
For the same peak power transmitter, the RADE waveform would have a mean signal power up to 7~dB higher than SSB at the receiver, leading to an additional 7~dB improvement.
This is however dependant on the SSB compressor employed (there is no standard PAPR for SSB), and for the RADE case assumes a transmitter that doesn't degrade the PAPR of the RADE signal.

The WER of FARGAN with clean features is very close to that of clean speech, thus our use of classical features and choice of vocoder is not a limiting factor.

\subsection{Complexity}

The encoder and decoder each require 1 MB of read-only storage (1M weights with 8-bit uniform quantization included in training to facilitate SIMD support) and require 32~MMACs for real-time operation.
Both easily operate in real time on a laptop using an unoptimised PyTorch implementation.
The overall complexity of the complete system is dominated by the FARGAN vocoder's 300~MMACs~\cite{valin2024low}.

\subsection{Over the Air Demonstration}
\label{sec:rade_ota}

To test RADE over real world HF channels licensed Amateur Radio operators from around the world were invited to record a 10 second input sample of their own voice. 
This was converted to a RADE waveform sample, and concatenated with a Hilbert-compressed version of the same input sample to emulate an analog SSB signal with a 6\nobreakdash-8~dB PAPR.
Participants then played the concatenated sample through their SSB transmitters over real world HF Radio channels to remote KiwiSDR receivers of their choice (KiwiSDRs are SSB radio receivers that are connected to the public Internet). 
The received off air signal was processed to obtain a file of SSB and RADE output speech samples, and an estimate of channel SNR. 
The use of stored files enabled the same Tx signal to be transmitted to different KiwiSDR receivers under different channel conditions, and different transmit power levels.
Tests were performed in English, Japanese, Cantonese, and German, over a variety of HF channels of up to 14,000~km (e.g. direct transmission from North America to Australia). Speech samples are available at~\cite{sep24radedemo}. Source code is available on GitHub~\cite{rade_github} in the \texttt{waspaa\_2025} branch.

\section{Conclusion}
\label{sec:conclusion}

We have combined a ML vocoder, ML autoencoder and classical DSP OFDM to build a system capable of sending speech over HF radio channels. 
It is robust to AWGN and multipath channel impairments, and the transmit signal has a PAPR of less than 1~dB. 
Unusually for HF speech systems, the audio bandwidth is 8000~Hz, despite requiring just 1500~Hz of RF bandwidth. 
Our ASR simulation and real world demonstration show performance significantly exceeding that of analog SSB at the same SNR.
Unlike classical DSP HF systems, speech quality improves gradually with channel SNR without any mode switching.
Interestingly, our system also shows robustness to channel impairments we did not train for, e.g. impulse noise.

The experimental HF OTA results demonstrate surprisingly good performance on multipath channels where the period of the fading (100s of ms) is large compared to the 40~ms analysis window of the autoencoder. 
To overcome fading with classical DSP requires an interleaver of several times the fading period (e.g. 1000-2000~ms) which introduces significantly more algorithmic delay than our system for a similar level of robustness.
 
The OFDM frame design contains three latent vectors $\mathbf{z}$, so introduces an algorithmic delay of 120~ms.
This is comparable to other digital PTT radio systems, e.g. P.25~\cite{project25}.

\bibliographystyle{IEEEtran}
\bibliography{2024_rade_hf_refs}

\end{document}